\documentclass[11pt]{article}
\usepackage{amssymb,amsmath,amsfonts}
\usepackage{graphicx}
\usepackage{graphics}
\usepackage{eepic,epsfig}
\usepackage{dcolumn}

\begin{document}

\title{Synchrotron radiation from a charge circulating \\
around a cylinder with negative permittivity}
\author{A.A. Saharian$^{1,2}$\thanks{%
E-mail: saharian@ysu.am}, A.S. Kotanjyan$^{2}$, L.Sh. Grigoryan$^{1}$, H.F. Khachatryan$^{1}$%
, \\ V.Kh. Kotanjyan$^{1}$ \vspace{0.3cm} \\
\textit{$^{1}$Institute of Applied Problems in Physics, }\\
\textit{25 Nersessian Street, 0014 Yerevan, Armenia} \vspace{0.3cm}\\
\textit{$^{2}$Department of Physics, Yerevan State University,}\\
\textit{1 Alex Manoogian Street, 0025 Yerevan, Armenia}\vspace{0.3cm}}
\maketitle

\begin{abstract}
We investigate the radiation from a charged particle rotating around a
dielectric cylinder with a negative real part of dielectric permittivity.
For the general case of frequency dispersion in dielectric permittivity,
expressions are derived for the electric and magnetic fields and for the
angular density of the radiation intensity on a given harmonic. Compared
with the case of a cylinder with positive real part of the permittivity, new
interesting features arise in the nonrelativistic limit and for the
radiation at small angles with respect to the cylinder axis. Another feature
is the appearance of sharp narrow peaks in the angular density of the
radiation intensity for large harmonics. We analytically estimate the
height, width and the location of these peaks. The influence of the
imaginary part of dielectric permittivity on the characteristics of the
peaks is discussed. The analytical results are illustrated by numerical
examples. We show that the radiation intensity on a given harmonic,
integrated over the angles, can be essentially amplified by the presence of
the cylinder.
\end{abstract}

\section{Introduction}

\label{sec:oscint}

It is well-known that the presence of medium may essentially enhance the
output power of various types of radiation processes. In addition, the
polarization of media by charged particles gives rise new types of
radiations, such as Cherenkov, diffraction and transition radiations. New
interesting features in the radiation processes appear in spectral ranges
where the real part of the dielectric permittivity of material becomes
negative (for general consideration concerning the existence of a negative
dielectric constant see Ref. \cite{Dolg81}). The metals provide an example
of this kind of material. Due to relatively large densities of free carriers
they exhibit a negative permittivity from the visible to microwave
frequencies. Another example of material with negative real part of the
dielectric permittivity is provided by doped semiconductors \cite%
{West10,Naik10,Mitt13}. Compared to the metals, the doped semiconductors can
exhibit very small losses at infrared and longer wavelengths and the
corresponding plasma frequency can be controlled by tuning the free carrier
densities.

At interfaces between two media with positive and negative real parts of the
dielectric permittivity new types of surface waves arise called surface
polaritons. In particular, the surface plasmon polaritons (SPPs) have found
a wide range of applications including surface imaging, surface-enhanced
Raman spectroscopy, data storage, biosensors, plasmonic waveguides,
photovoltaics, various types of light-emitting devices, plasmonic solar
cells, etc. Refs. \cite{Agra82}-\cite{Enoc12}. SPPs are evanescent
electromagnetic waves propagating along a metal-dielectric interface as a
result of collective oscillations of electron subsystem coupled to
electromagnetic field. Several techniques are available for generation of
surface polaritons. In particular, the surface polaritons can be excited by
electron beams moving parallel or perpendicularly to the interface (see, for
instance, Refs. \cite{Raet88}-\cite{Gong14} and references therein). The
radiation of surface polaritons by a charged particle rotating around a
cylindrical dielectric waveguide recently has been considered in Ref. \cite%
{Kota19}. It has been shown that the radiation intensity for surface
polaritons of a given harmonic can be essentially larger than that for
guiding modes of the cylinder (the radiation on guiding modes for a cylinder
with positive dielectric permittivity is investigated in Refs. \cite%
{Kota18,Saha19}).

For a charge rotating around a cylindrical rode, in addition to the
radiation of surface-type modes (surface polaritons) and guided modes, where
will be radiation propagating at large distances from the cylinder. That
corresponds to the synchrotron radiation \cite{Soko86,Hofm04} influenced by
the presence of the cylinder. For a cylinder with positive dielectric
permittivity, in Refs. \cite{Grig95,Kota00} it has been shown that for the
rotation orbit close to the cylinder surface the influence of the cylinder
on the angular distribution of the radiation intensity on a given harmonic
can be essential. In particular, under the Cherenkov condition for the
velocity of the charge image on the cylinder surface and for the cylinder
dielectric permittivity, strong narrow peaks may appear in that distribution
(similar features for the radiation from a charge circulating around a
dielectric ball were discussed in Refs. \cite{Grig98,Grig06}). The
corresponding results for a charge moving along a helical trajectory around
a cylinder are presented in Ref. \cite{Saha09,Kota13}. In the present paper
we consider the features for the radiation from a charge rotating around a
dielectric cylinder in the frequency range with negative real part of
dielectric permittivity of the cylinder material.

The organization of the paper is as follows. In the next section we describe
the problem setup and present the electric and magnetic fields in the region
outside the cylinder. The spectral-angular distribution of the radiation
intensity is investigated in Sect. \ref{sec:radiation}. The features of the
radiation are discussed in various asymptotic regions of the parameters.
Numerical examples illustrating the general results are presented in Sect. %
\ref{sec:Numer}. Section \ref{sec:Conc} concludes the main results of the
paper.

\section{Problem setup and the electromagnetic fields}

\label{sec:oscfields}

\begin{figure}[tbph]
\begin{center}
\epsfig{figure=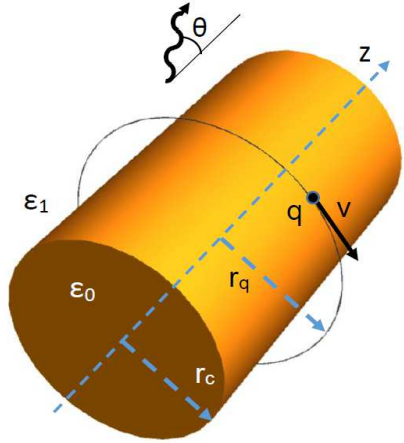,width=5.5cm,height=6cm}
\end{center}
\caption{Point charge rotating around a cylinder.}
\label{fig1}
\end{figure}

We consider a particle with charge $q$ moving along a circular trajectory of
radius $r_{q}$ around a cylinder with radius $r_{c}$ and with dielectric
permittivity $\varepsilon _{0}$. The radii of the particle trajectory and of
the cylinder will be denoted by $r_{q}$ and $r_{c}$, respectively, and it
will be assumed that the system is embedded in a homogeneous medium with
dielectric permittivity $\varepsilon _{1}$. We will consider the general
case of frequency dependent complex permittivity $\varepsilon
_{0}=\varepsilon _{0}^{\prime }(\omega )+i\varepsilon _{0}^{\prime \prime
}(\omega )$. In accordance with the problem symmetry we will use the
cylindrical coordinate system $(r,\phi ,z)$ with the axis $z$ along the
cylinder axis (the geometry of the problem is depicted in Fig. \ref{fig1}).
For the components of the current density in that coordinates one has
\begin{equation}
j_{l}=\frac{q}{r}v\delta _{l\phi }\delta (r-r_{q})\delta (\phi -\omega
_{0}t)\delta (z),  \label{jl}
\end{equation}%
where $l=r,\phi ,z$, $v$ is the particle velocity and $\omega _{0}=v/r_{q}$
is the angular velocity.

Here we are interested in the radiation at large distances from the cylinder
in the frequency range where the real part of the dielectric permittivity $%
\varepsilon _{0}$ of the cylinder is negative, $\varepsilon _{0}^{\prime }<0$%
. We assume that the host medium is transparent and the corresponding
dielectric permittivity $\varepsilon _{1}$ is real and positive. Under these
conditions there are two types of radiations: the radiation propagating at
large distances from the cylinder and the radiation of surface-type modes
localized near the cylinder surface. For a metallic cylinder the latter
corresponds to SPPs, whereas the first type of radiation corresponds to the
synchrotron radiation influenced by the cylinder. As it will be shown below,
that influence can be essential. The frequency range of negative $%
\varepsilon _{0}^{\prime }$ is also of special interest for the optics of
small particles \cite{Bohr83}. Among the simplest models describing the
dispersion of dielectric permittivity is the generalized Drude model where
the conduction electrons are considered as a free-electron gas (see, for
instance, Refs. \cite{West10,Bohr83}). The corresponding frequency
dependence is given by the expression%
\begin{equation}
\varepsilon _{0}\left( \omega \right) =\varepsilon _{\mathrm{b}}-\frac{%
\omega _{p}^{2}}{\omega \left( \omega +i\gamma \right) },  \label{Drude}
\end{equation}%
where $\omega _{p}$ is the plasma frequency and $1/\gamma $ is the mean
relaxation time of conduction electrons. The part $\varepsilon _{\mathrm{b}}$
is related to the contribution of bound-electrons and can be described by
the Lorentz oscillator model. In some frequency ranges it can be
approximated as constant. The plasmonic effects in metals and doped
semiconductors are mainly discussed on the base of the model (\ref{Drude}).
The doped semiconductors with low electron density are used to bring the
plasma frequency down to THz range. Another approach is based on the use of
appropriately designed metamaterials.

First we consider the electromagnetic fields created by the current density (%
\ref{jl}). For the corresponding electric, $\mathbf{E}(\mathbf{r},t)$, and
magnetic, $\mathbf{H}(\mathbf{r},t)$, fields one has the following Fourier
expansion
\begin{equation}
\mathbf{F}(\mathbf{r},t)=\sum_{n=-\infty }^{\infty }e^{in\phi -i\omega
_{n}t)}\int_{-\infty }^{\infty }dk_{z}\,e^{ik_{z}z}\mathbf{F}_{n}(k_{z},r),
\label{F}
\end{equation}%
with $\omega _{n}=n\omega _{0}$ and $\mathbf{F}=\mathbf{E},\mathbf{H}$. By
taking into account that $\mathbf{F}_{-n}(k_{z},r)=\mathbf{F}_{n}^{\ast
}(-k_{z},r)$, where the star means the complex conjugate, in the discussion
below for the Fourier components $\mathbf{F}_{n}(k_{z},r)$ we will assume
that $n>0$. Note that the $n=0$ term in Eq. (\ref{F}) is time independent
and will not contribute to the radiation fields. The Fourier components of
the fields can be found by using the Green function from Ref. \cite{Grig95}
in a way similar to that presented in Refs. \cite{Grig95,Kota00,Saha09} for
a cylinder with positive dielectric permittivity and in what follows we will
omit the details.

Denoting the cylindrical components as $F_{nl}(k_{z},r)$, with $l=r,\phi ,z$%
, and for simplicity of the presentation omitting the arguments $(k_{z},r)$,
for the magnetic field in the region $r>r_{q}$ one gets
\begin{eqnarray}
H_{nl} &=&\frac{i^{2-\sigma _{l}}qvk_{z}}{2\pi c}\sum_{p=\pm 1}p^{\sigma
_{l}-1}B_{n,p}H_{n+p}(\lambda r),\;l=r,\phi ,  \notag \\
H_{nz} &=&-\frac{qv\lambda }{2\pi c}\sum_{p=\pm 1}pB_{n,p}H_{n}(\lambda r),
\label{Hnl}
\end{eqnarray}%
where $\sigma _{r}=1$, $\sigma _{\phi }=2$, $H_{n}(x)\equiv H_{n}^{(1)}(x)$
is the Hankel function of the first kind and
\begin{equation}
\lambda ^{2}=\omega _{n}^{2}\varepsilon _{1}/c^{2}-k_{z}^{2},\;\omega
_{n}=n\omega _{0}.  \label{lamb}
\end{equation}%
The functions $B_{n,p}$ are defined by the expression
\begin{equation}
B_{n,p}=\frac{\pi }{2i}J_{n+p}(\lambda r_{q})-\frac{\pi }{2i}H_{n+p}(\lambda
r_{q})\frac{W_{n+p}^{J}}{W_{n+p}^{H}}+p\frac{\eta I_{n}(\eta r_{c})}{%
2r_{c}\alpha _{n}}\frac{I_{n+p}(\eta r_{c})}{W_{n+p}^{H}}\sum_{l=\pm 1}l%
\frac{H_{n+l}(\lambda r_{q})}{W_{n+l}^{H}},  \label{Bnp}
\end{equation}%
with $J_{n}(x)\ $and $I_{n}(x)$ being the Bessel and modified Bessel
functions, and
\begin{equation}
\eta ^{2}=k_{z}^{2}-\omega _{n}^{2}\varepsilon _{0}/c^{2}.  \label{eta1}
\end{equation}%
The other notations in Eq. (\ref{Bnp}) are defined as
\begin{equation}
\alpha _{n}=\frac{\varepsilon _{0}}{\varepsilon _{1}-\varepsilon _{0}}-\frac{%
\eta I_{n}(\eta r_{c})}{2}\sum_{l=\pm 1}\frac{H_{n+l}(\lambda r_{c})}{%
W_{n+l}^{H}},  \label{alfn}
\end{equation}%
and
\begin{equation}
W_{n+p}^{F}=p\lambda I_{n+p}(\eta r_{c})F_{n}(\lambda r_{c})-\eta
F_{n+p}(\lambda r_{c})I_{n}(\eta r_{c}),  \label{WF}
\end{equation}%
for the Bessel and Hankel functions $F_{\nu }(x)=J_{\nu }(x),H_{\nu }(x)$.
The Fourier components for the electric field are obtained from Eq. (\ref%
{Hnl}) by using the Maxwell equation $\partial \mathbf{E}/\partial
t=(c/\varepsilon _{1})\nabla \times \mathbf{H}$. In the region $r>r_{q}$ one
gets
\begin{eqnarray}
E_{nl} &=&\frac{i^{1-\sigma _{l}}qv}{4\pi \omega _{n}\varepsilon _{1}}%
\sum_{p=\pm 1}p^{\sigma _{l}}\left[ \left( \frac{\omega _{n}^{2}}{c^{2}}%
\varepsilon _{1}+k_{z}^{2}\right) B_{n,p}-\lambda ^{2}B_{n,-p}\right]
H_{n+p}(\lambda r),  \notag \\
E_{nz} &=&\frac{iqv\lambda k_{z}}{2\pi \omega _{n}\varepsilon _{1}}%
\sum_{p=\pm 1}B_{n,p}H_{n}(\lambda r).  \label{Enl}
\end{eqnarray}%
where $l=r,\phi $.

The parts in the fields (\ref{Hnl}) and (\ref{Enl}) with the first term in
the right-hand side of Eq. (\ref{Bnp}) do not depend on $\varepsilon _{0}$
and correspond to the fields for a charge rotating in a homogeneous medium
with permittivity $\varepsilon _{1}$ when the cylinder is absent. The
corresponding Fourier components will be denoted by $F_{nl}^{(0)}(k_{z},r)$,
$F=H,E$. They are given by expressions which are obtained from Eqs. (\ref%
{Hnl}) and (\ref{Enl}) making the replacement $B_{n,p}\rightarrow
B_{n,p}^{(0)}$ with
\begin{equation}
B_{n,p}^{(0)}=\frac{\pi }{2i}J_{n+p}(\lambda r_{q}).  \label{Bnp0}
\end{equation}%
The parts in the fields with the last two terms in Eq. (\ref{Bnp}) are
induced by the cylinder. The expressions for the Fourier components $%
F_{nl}(k_{z},r)$ in the region $r_{c}<r<r_{q}$ are obtained from Eqs. (\ref%
{Hnl}) and (\ref{Enl}) making the replacements $J\rightleftarrows H$ of the
Bessel and Hankel functions in the parts corresponding to the fields $%
F_{nl}^{(0)}(k_{z},r)$ in a homogeneous medium. Note that the radial
dependence of the Fourier components $F_{nl}(k_{z},r)$ inside the cylinder
is described by the functions $I_{n}(\eta r)$ and $I_{n\pm 1}(\eta r)$.

The radiation fields at large distances from the cylinder correspond to the
integration range $k_{z}^{2}<\omega _{n}^{2}\varepsilon _{1}/c^{2}$ in Eq. (%
\ref{F}), where $\lambda $, defined by Eq. (\ref{lamb}), is real. For the
parts with the integration range $k_{z}^{2}>\omega _{n}^{2}\varepsilon
_{1}/c^{2}$ the quantity $\lambda $ is purely imaginary and the fields
induced by the cylinder depend on the radial coordinate through the
Macdonald functions $K_{n+p}(|\lambda |r)$. These parts are exponentially
suppressed at large distances from the cylinder and for a cylinder with a
negative real part of $\varepsilon _{0}$ they correspond to the surface-type
modes. Note that for the surface-type modes the allowed values of $k_{z}$
are the solutions of the equation $\alpha _{n}=0$. The radiation fields and
the radiation intensity for this type of modes have been discussed in Ref.
\cite{Kota19}. In what follows we are interested in the radiation at large
distances from the cylinder.

\section{Spectral and angular distribution of the radiation intensity}

\label{sec:radiation}

Having the electromagnetic fields we can investigate the intensity of the
radiation propagating in the exterior medium. As we have mentioned before,
for $\lambda ^{2}<0$ the corresponding Fourier components are exponentially
damped for large values $r$, and the radiation is present only under the
condition $\lambda ^{2}>0$. The average energy flux per unit time through
the cylindrical surface of radius $r>r_{c}$, coaxial with the dielectric
cylinder, is given by the Poynting vector:
\begin{equation}
I=\frac{c}{2T}\int_{0}^{T}dt\int_{-\infty }^{\infty }dz\,r\mathbf{n}%
_{r}\cdot \left[ \mathbf{E}\times \mathbf{H}\right] ,  \label{I1}
\end{equation}%
where $T=2\pi /\omega _{0}$ is the period of the charge rotation and $%
\mathbf{n}_{r}$ is the unit vector along the radial direction. Substituting
the Fourier expansions (\ref{F}) for the electric and magnetic fields, we
obtain
\begin{equation}
I=2c\pi r\,\mathrm{Re}\left\{ \sum_{n=1}^{\infty }{}\int_{\lambda
_{1}^{2}>0}dk_{z}\,[E_{n\phi }H_{nz}^{\ast }-E_{nz}H_{n\phi }^{\ast
}]\right\} ,  \label{I2}
\end{equation}%
where
\begin{eqnarray}
E_{n\phi }H_{nz}^{\ast }-E_{nz}H_{n\phi }^{\ast } &=&\frac{%
iq^{2}v^{2}\lambda }{4\pi ^{2}c\omega _{n}\varepsilon _{1}}\left[
k_{z}^{2}H_{n}(\lambda r)H_{n}^{\ast \prime }(\lambda r)\left\vert
B_{n,1}+B_{n,-1}\right\vert ^{2}\right.  \notag \\
&&\left. -\omega _{n}^{2}c^{-2}\varepsilon _{1}H_{n}^{\prime }(\lambda
r)H_{n}^{\ast }(\lambda r)\left\vert B_{n,1}-B_{n,-1}\right\vert ^{2}\right]
.  \label{EH}
\end{eqnarray}%
and the prime means the derivative with respect to the argument of the
function. At large distances, assuming that $\lambda r\gg 1$, we use in Eq. (%
\ref{EH}) the asymptotic expressions for the Hankel functions for large
arguments.

Passing in Eq. (\ref{I2}) to a new integration variable $\theta $, $0\leq
\theta \leq \pi $, in accordance with
\begin{equation*}
k_{z}=\frac{\omega _{n}}{c}\sqrt{\varepsilon _{1}}\cos \theta ,
\end{equation*}%
the energy flux is presented in the form
\begin{equation}
I=\sum_{n=1}^{\infty }\int d\Omega \,\frac{dI_{n}}{d\Omega },  \label{I3}
\end{equation}%
where $d\Omega =\sin \theta d\theta d\phi $ is the solid angle element, and
\begin{equation}
\frac{dI_{n}}{d\Omega }=\frac{q^{2}v^{2}\omega _{n}^{2}\sqrt{\varepsilon _{1}%
}}{2\pi ^{3}c^{3}}\left[ \left\vert B_{n,1}-B_{n,-1}\right\vert
^{2}+\left\vert B_{n,1}+B_{n,-1}\right\vert ^{2}\cos ^{2}\theta \right] .
\label{In4}
\end{equation}%
The angular variable $\theta $ is the angle between the wave vector of the
radiated photon and the cylinder axis (axis $z$). Eq. (\ref{I3}) presents
the average power radiated by the charge on a given frequency $\omega
=\omega _{n}$ into a unit solid angle. In Eq. (\ref{In4}), the functions $%
B_{n,p}$ are given by Eq. (\ref{Bnp}), where now we should take%
\begin{equation}
\lambda =\frac{\omega _{n}}{c}\sqrt{\varepsilon _{1}}\sin \theta ,
\label{lam1}
\end{equation}%
and%
\begin{equation}
\eta =\frac{\omega _{n}}{c}\sqrt{\varepsilon _{1}\cos ^{2}\theta
-\varepsilon _{0}}.  \label{eta}
\end{equation}%
Note that Eq. (\ref{In4}) is valid for a general case of dispersion $%
\varepsilon _{0}=\varepsilon _{0}(\omega _{n})$ of the cylinder material
with complex permittivity. The expression for the angular density of the
radiation intensity from a charge rotating in a homogeneous medium with
permittivity $\varepsilon _{1}$ (denoted here as $dI_{n}^{(0)}/d\Omega $) is
obtained from Eq. (\ref{In4}) by the replacement $B_{n,p}\rightarrow
B_{n,p}^{(0)}$, where $B_{n,p}^{(0)}$ is given by Eq. (\ref{Bnp0}). By using
the recurrence relations for the Bessel function one gets (see, for
instance, Refs. \cite{Soko86,Zrel70})%
\begin{equation}
\frac{dI_{n}^{(0)}}{d\Omega }=\frac{q^{2}\omega _{n}^{2}}{2\pi c\sqrt{%
\varepsilon _{1}}}\left[ \beta _{1}^{2}J_{n}^{\prime 2}(n\beta _{1}\sin
\theta )+\cot ^{2}\theta J_{n}^{2}(n\beta _{1}\sin \theta )\right] ,
\label{In0}
\end{equation}%
with the notation $\beta _{1}=v\sqrt{\varepsilon _{1}}/c$. For relativistic
particles, $v\approx c$, the radiation frequency is determined by the
harmonic number and by the radius of the rotation orbit $r_{q}$. For $r_{q}$
of the order of 1 mm and for the harmonics of the order 10, the radiation
corresponds to the THz frequency range (for configurations of external
fields generating that type of circular motion for electron see, for
example, Ref. \cite{Saha07}).

Let us consider the behavior of the radiation intensity in some asymptotic
regions of the parameters. For a nonrelativistic charge, assuming that $%
n\beta _{1}\ll 1$, in Eq. (\ref{Bnp}) we can use the asymptotic expressions
of cylindrical functions for small arguments \cite{Abra72}. To the leading
order this gives%
\begin{equation}
B_{n,p}\approx \frac{\pi }{2i}\frac{\left( \lambda r_{q}/2\right) ^{n+p}}{%
\Gamma \left( n+p+1\right) }\left[ 1-\frac{p}{2\alpha _{n}}\left(
r_{c}/r_{q}\right) ^{2n+p+1}\right] ,  \label{Bnp1}
\end{equation}%
where%
\begin{equation}
\alpha _{n}\approx \frac{1}{2}\frac{\varepsilon _{1}+\varepsilon _{0}}{%
\varepsilon _{1}-\varepsilon _{0}}+\frac{r_{c}^{2}}{8n}\left( \frac{\lambda
^{2}}{n+1}+\frac{\eta ^{2}}{n-1}\right) ,  \label{alfn3}
\end{equation}%
for $n\geq 2$ and
\begin{equation}
\alpha _{1}\approx \frac{1}{2}\frac{\varepsilon _{1}+\varepsilon _{0}}{%
\varepsilon _{1}-\varepsilon _{0}}+\frac{r_{c}^{2}}{16}\left[ -4\eta ^{2}\ln
(\lambda r_{c}/2)+\lambda ^{2}\right] .  \label{alfn4}
\end{equation}%
Here, $\lambda $ and $\eta $ are given by the expressions (\ref{lam1}) and (%
\ref{eta}). In the expression for $\alpha _{n}$ we have kept the
next-to-leading order terms in order to consider the behavior of the
radiation intensity in the frequency range where $\varepsilon _{0}^{\prime }$
is close to $-\varepsilon _{1}$ and for small values of the imaginary part $%
\varepsilon _{0}^{\prime \prime }$ the leading term in the expansion is
small. As seen, for a given $n$, the dominant contribution to the radiation
intensity (\ref{In4}) comes from the terms with $B_{n,-1}$ and we get
\begin{equation}
\frac{dI_{n}}{d\Omega }\approx \frac{2q^{2}c\left( n\beta _{1}/2\right)
^{2n+2}}{\pi (n!)^{2}\varepsilon _{1}^{3/2}r_{q}^{2}}\left\vert 1+\frac{%
\left( r_{c}/r_{q}\right) ^{2n}}{2\alpha _{n}}\right\vert ^{2}\left( 1+\cos
^{2}\theta \right) \sin ^{2n-2}\theta ,  \label{In5}
\end{equation}%
with $\alpha _{n}$ from Eqs. (\ref{alfn3}) and (\ref{alfn4}). Compared to
the radiation on the main harmonic $n=1$, the radiation on higher harmonics $%
n\geq 2$ is suppressed by the factor $\beta _{1}^{2n-2}$. The part in Eq. (%
\ref{In5}) with the first term in the expression under the absolute sign
corresponds to the radiation in a homogeneous medium with permittivity $%
\varepsilon _{1}$.

For positive $\varepsilon _{0}^{\prime }$ we can keep the leading order
terms and in Eq. (\ref{In5}) $\alpha _{n}\approx \left( \varepsilon
_{1}+\varepsilon _{0}\right) /[2(\varepsilon _{1}-\varepsilon _{0})]$. In
this case, in the expression under the sign of modulus in eq. (\ref{In5})
the contribution of the term induced by the cylinder is smaller than the one
corresponding to the radiation in a homogeneous medium and the radiation
intensity behaves as $\beta _{1}^{2n+2}$. A new qualitatively different
feature arises for negative values of $\varepsilon _{0}^{\prime }$ and for
small $\varepsilon _{0}^{\prime \prime }$. In this case, under the
assumption $|\varepsilon _{1}+\varepsilon _{0}^{\prime }|,\varepsilon
_{0}^{\prime \prime }\lesssim \beta _{1}^{2}$, we see that $\alpha
_{n}\propto \beta _{1}^{2}$ for $n\geq 2$ and $\alpha _{1}\propto \beta
_{1}^{2}\ln (\beta _{1})$. Now, the dominant contribution to the radiation
intensity (\ref{In5}) comes from the part induced by the cylinder and one
has $dI_{n}/d\Omega \propto \beta _{1}^{2(n-1)}$ for $n\geq 2$ and $%
dI_{1}/d\Omega \propto 1/\ln ^{2}(\beta _{1})$.

Now let us consider the radiation intensity for small values of the angle $%
\theta $ and for fixed values of the other parameters. In this limit,
assuming that
\begin{equation}
\lambda r_{q}=n\beta _{1}\sin \theta \ll 1,  \label{smang}
\end{equation}%
for cylinder functions having in their arguments $\lambda r_{q}$ and $%
\lambda r_{c}$ we use the asymptotic expressions for small arguments. In
particular, for the function $\alpha _{n}$ to the leading order we get
\begin{eqnarray}
\alpha _{1} &\approx &\frac{\varepsilon _{1}}{\varepsilon _{1}-\varepsilon
_{0}},  \notag \\
\alpha _{n} &\approx &\frac{1}{2}\frac{\varepsilon _{1}+\varepsilon _{0}}{%
\varepsilon _{1}-\varepsilon _{0}}+\left[ (n-1)\frac{2I_{n-1}(\eta r_{c})}{%
\eta r_{c}I_{n}(\eta r_{c})}+1\right] ^{-1},  \label{alfnSmang}
\end{eqnarray}%
where $n\geq 2$ and $\eta r_{c}\approx n(\omega _{0}r_{c}/c)\sqrt{%
\varepsilon _{1}-\varepsilon _{0}}$. In the limit $\theta \rightarrow 0$,
the angular density of the radiation intensity tends to zero as $%
dI_{n}/d\Omega \propto 1/\ln (\sin \theta )$ for $n=1$ and as $%
dI_{n}/d\Omega \propto \sin ^{2n-2}\theta $ for $n\geq 2$. For small angles $%
\theta $, assuming that $\varepsilon _{0}^{\prime \prime }\lesssim \sin
\theta $, in the leading terms (\ref{alfnSmang}) we can replace $\varepsilon
_{0}$ by its real part $\varepsilon _{0}^{\prime }$. In this case, an
interesting situation arises for $\varepsilon _{1}+\varepsilon _{0}^{\prime
}<0$ and $n\geq 2$ when the leading term in (\ref{alfnSmang}) (with the
replacement $\varepsilon _{0}\rightarrow \varepsilon _{0}^{\prime }$) may
become zero. For given $n$ and $\omega _{0}r_{c}\sqrt{\varepsilon _{1}}/c$,
that condition can be considered as an equation for the corresponding value
of the ratio $\varepsilon _{0}/\varepsilon _{1}$. In this case the
next-to-leading term should be kept in the asymptotic expansion of $\alpha
_{n}$ over $\sin \theta $ and $\varepsilon _{0}^{\prime \prime }$. Near that
specified value of $\varepsilon _{0}/\varepsilon _{1}$ the decay of $%
dI_{n}/d\Omega $ in the limit $\theta \rightarrow 0$ is slower.

Another new qualitative feature in the radiation intensity, induced by the
cylinder, is the possibility for the appearance of strong narrow peaks for
large values of the radiation harmonic at specific values of the angle $%
\theta $. For a transparent cylinder with a positive dielectric
permittivity, this feature has been discussed in Refs. \cite{Kota00,Saha09}.
Here the characteristics of the peaks will be considered in the frequency
range with $\varepsilon _{0}^{\prime }<0$.

We start the discussion with the case of a transparent cylinder when the
imaginary part of $\varepsilon _{0}$ can be neglected and $\varepsilon
_{0}=\varepsilon _{0}^{\prime }$ is real. The peaks we are going to consider
arise under the condition $\lambda r_{c}<n$ which corresponds to the angular
range $\sin \theta <1/\beta _{c}$, where $\beta _{c}=v_{c}\sqrt{\varepsilon
_{1}}/c$ and $v_{c}=\omega _{0}r_{c}$ is the velocity of the charge image on
the cylinder surface. In this range, by using the asymptotic expressions for
the functions $J_{n}(ny)$ and $Y_{n}(ny)$ for $n\gg 1$(see, for instance,
Ref. \cite{Abra72}), we can show that%
\begin{eqnarray}
J_{n+p}(ny) &\sim &\frac{e^{-n\zeta (y)}}{\sqrt{2\pi n}y^{|p|}}\frac{1-p%
\sqrt{1-y^{2}}}{(1-y^{2})^{1/4}},  \notag \\
Y_{n+p}(ny) &\sim &-\frac{2e^{n\zeta (y)}}{\sqrt{2\pi n}y^{|p|}}\frac{1+p%
\sqrt{1-y^{2}}}{(1-y^{2})^{1/4}},  \label{Yas}
\end{eqnarray}%
where $p=0,\pm 1$, $0<y<1$ and
\begin{equation}
\zeta (y)=\ln \frac{1+\sqrt{1-y^{2}}}{y}-\sqrt{1-y^{2}}.  \label{dset}
\end{equation}%
As it can be seen from Eqs. (\ref{Bnp}) and (\ref{alfn}), the modified
Bessel functions $I_{n}(\eta r_{c})$ and $I_{n+l}(\eta r_{c})$, $l=\pm 1$,
appear in the expression of the radiation intensity in the form of the ratio
$I_{n+l}(\eta r_{c})/I_{n}(\eta r_{c})$. For the latter, the asymptotic
expression for large values of $n$ reads%
\begin{equation}
\frac{I_{n+l}(nu)}{I_{n}(nu)}\sim \frac{\sqrt{1+u^{2}}-l}{u},\;l=\pm 1.
\label{Inas}
\end{equation}%
The mathematical reason for the appearance of the above mentioned peaks is
the exponential suppression of the ratio $J_{n+p}(ny)/Y_{n+p}(ny)\propto
e^{-2n\zeta (y)}$ for large $n$.

For the further analysis of the peaks in the angular distribution of the
radiation intensity it is convenient to rewrite the function $\alpha _{n}$
in the form%
\begin{equation}
\alpha _{n}=\frac{\varepsilon _{0}}{\varepsilon _{1}-\varepsilon _{0}}+\frac{%
1}{2}\sum_{l=\pm 1}\left[ 1-l\frac{\lambda I_{n+l}(\eta r_{c})H_{n}(\lambda
r_{c})}{\eta I_{n}(\eta r_{c})H_{n+l}(\lambda r_{c})}\right] ^{-1}.
\label{alfn5}
\end{equation}%
For large values of $n$ and under the condition $\lambda r_{c}<n$, we expand
Eq. (\ref{alfn5}) with respect to small ratio $J_{n+l}(\lambda
r_{c})/Y_{n+l}(\lambda r_{c})$ with $l=0,\pm 1$. The leading term is
obtained from Eq. (\ref{alfn5}) by the replacement $H_{n+l}(\lambda
r_{c})\rightarrow Y_{n+l}(\lambda r_{c})$:%
\begin{equation}
\alpha _{n}^{(0)}=\frac{\varepsilon _{0}}{\varepsilon _{1}-\varepsilon _{0}}+%
\frac{1}{2}\sum_{l=\pm 1}\left[ 1-l\frac{\lambda I_{n+l}(\eta
r_{c})Y_{n}(\lambda r_{c})}{\eta I_{n}(\eta r_{c})Y_{n+l}(\lambda r_{c})}%
\right] ^{-1}.  \label{alfn50}
\end{equation}%
The peaks arise near the angles $\theta $ for which this leading term
becomes zero: $\alpha _{n}^{(0)}=0$. Near these angles, keeping the
next-to-leading order term one gets%
\begin{equation}
\alpha _{n}\approx \frac{i\eta }{\pi r_{c}Y_{n}^{2}(\lambda r_{c})}%
\sum_{l=\pm 1}\frac{I_{n+l}(\eta r_{c})}{I_{n}(\eta r_{c})}\left[ \eta \frac{%
Y_{n+l}(\lambda r_{c})}{Y_{n}(\lambda r_{c})}-l\lambda \frac{I_{n+l}(\eta
r_{c})}{I_{n}(\eta r_{c})}\right] ^{-2},  \label{alfn6}
\end{equation}%
and, hence, for $n\gg 1$ in that range we have $\alpha _{n}\propto
e^{-2n\zeta (\lambda r_{c}/n)}$.

Now let us estimate the contribution of separate terms in $B_{n,p}$, defined
by Eq. (\ref{Bnp}), for large $n$. The contribution of the first term in the
right-hand side of Eq. (\ref{Bnp}) (corresponding to the radiation in a
homogeneous medium) is suppressed by the factor $e^{-n\zeta (\lambda
r_{q}/n)}$. The contribution of the second term is suppressed by $%
e^{-n[2\zeta (\lambda r_{c}/n)-\zeta (\lambda r_{q}/n)]}$. By taking into
account that the function $\zeta (y)$ is monotonically decreasing for $y>0$
we have $\zeta (\lambda r_{q}/n)<\zeta (\lambda r_{c}/n)$. Consequently, the
suppression of the second term is stronger than that for the first one (by
the relative factor $e^{-2n[\zeta (\lambda r_{c}/n)-\zeta (\lambda
r_{q}/n)]} $). If the angles $\theta $ are not close to the ones determined
by the zeros of $\alpha _{n}^{(0)}$, the function $\alpha _{n}$ is of the
order of one and the contribution of the third term in the right-hand side
of Eq. (\ref{Bnp}) is of the same order as that for the second term. In this
case the contribution of the first term in Eq. (\ref{Bnp}) dominates and the
radiation intensity is close to the one for a charge rotating in a
homogeneous medium. The situation is essentially different for the radiation
angles close to the ones determined by the zeros of $\alpha _{n}^{(0)}$. For
these angles one has $\alpha _{n}\propto e^{-2n\zeta (\lambda r_{c}/n)}$ and
the contribution of the last term in Eq. (\ref{Bnp}) is of the order $%
e^{n\zeta (\lambda r_{q}/n)}$. Hence, the angular density of the radiation
intensity for the peak at $\theta =\theta _{p}$ is proportional to $%
e^{2n\zeta (\beta _{1}\sin \theta _{p})}$. For the peak in the region $z>0$ (%
$0<\theta _{p}<\pi /2$) the height increases with decreasing $\theta _{p}$.
The angular width $\Delta \theta _{p}$ of the peak can be estimated in a way
similar to that given in Ref. \cite{Saha09} and is of the order $\exp
[-2n\zeta (\beta _{1}(r_{c}/r_{q})\beta _{1}\sin \theta _{p})]$. As seen,
with increasing height of the peak the corresponding width decreases. In the
next section this features will be illustrated by numerical examples.

\section{Numerical examples}

\label{sec:Numer}

\begin{figure}[tbph]
\begin{center}
\epsfig{figure=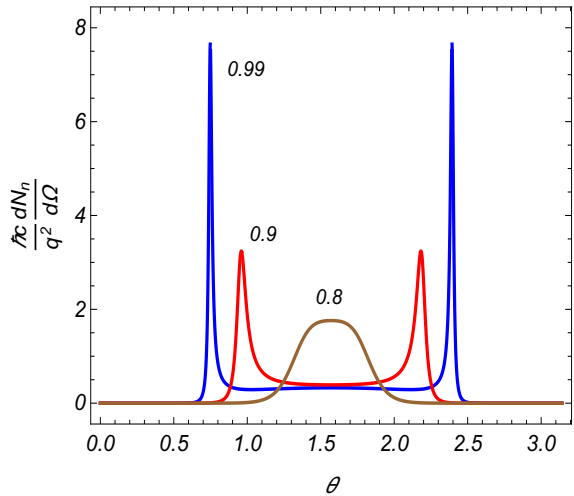,width=7.5cm,height=6.5cm}
\end{center}
\caption{The angular density of the number of the radiated quanta per
rotation period versus the radiation angle $\protect\theta $ for a cylinder
with permittivity $\protect\varepsilon _{0}=-3$ in the vacuum. The numbers
near the curves correspond to the values of the ratio $v/c$ and the graphs
are plotted for $r_{c}/r_{q}=0.95$, $n=10$.}
\label{fig2}
\end{figure}

In the numerical investigation of the spectral-angular distribution of the
radiation intensity at large distances from the cylinder we evaluate the
angular density for the number of the quanta radiated on a given harmonic
per period of the particle rotation:
\begin{equation}
\frac{dN_{n}}{d\Omega }=\frac{T}{\hbar \omega _{n}}\frac{dI_{n}}{d\Omega }.
\label{dNn}
\end{equation}%
In Fig. \ref{fig2} we display the dependence of this quantity (in units of $%
q^{2}/\hbar c$) on the radiation direction $\theta $ for a transparent
cylinder ($\varepsilon _{0}^{\prime \prime }=0$) with dielectric
permittivity $\varepsilon _{0}=-3$ in the vacuum ($\varepsilon _{1}=1$). The
graphs are plotted for $r_{c}/r_{q}=0.95$, $n=10$ and the numbers near the
curves correspond to the values of $v/c$. For the cases $v/c=0.9,0.99$ we
see the presence of the peaks described above analytically. For the angular
locations of the peaks one has $\theta \approx 0.96$ for $v/c=0.9$ and $%
\theta \approx 0.75$ for $v/c=0.99$. With increasing $v/c$, the angular
location of the peak in the region $0\leq \theta \leq \pi /2$ is shifted to
smaller angles. For the solution of the equation $\alpha _{n}^{(0)}=0$ in
the region $0<\theta <\pi /2$ one has $\theta \approx 0.965$ for $v/c=0.9$
and $\theta \approx 0.75$ for $v/c=0.99$. As seen, in agreement with the
analysis given above, these roots coincide with the locations of the peaks
in the radiation intensity with good accuracy.

\begin{figure}[tbph]
\begin{center}
\epsfig{figure=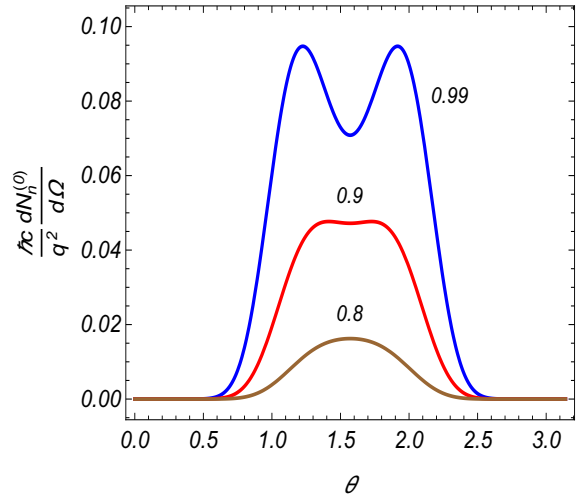,width=7.5cm,height=6.5cm}
\end{center}
\caption{The same as in Fig. \protect\ref{fig2} for the radiation of a
circulating charge in the vacuum ($\protect\varepsilon _{0}=\protect%
\varepsilon _{1}=1$). }
\label{fig3}
\end{figure}

In order to see the effect of the cylinder on the radiation intensity, in
Fig. \ref{fig3} we present the corresponding quantity for the radiation
intensity in the absence of cylinder ($\varepsilon _{0}=\varepsilon _{1}$), $%
dN_{n}^{(0)}/d\Omega $ (obtained from Eq. (\ref{In0}) in a way similar to
Eq. (\ref{dNn})), for the same values of the other parameters. As we see,
the presence of the cylinder may lead to an essential increase in the
angular density of the radiation intensity.

\begin{figure}[tbph]
\begin{center}
\epsfig{figure=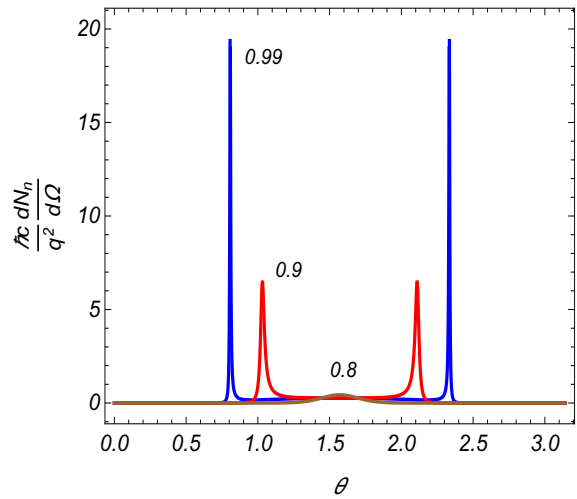,width=7.5cm,height=6.5cm}
\end{center}
\caption{The same as in Fig. \protect\ref{fig2} for $n=15$.}
\label{fig4}
\end{figure}

As it has been explained by the asymptotic analysis, the height of the
narrow peaks in the angular distribution of the radiation energy increases
with increasing $n$. This feature is seen in Fig. \ref{fig4}, where the
angular density of the number of radiated quanta is plotted versus $\theta $
for the harmonic $n=15$. The values of the other parameters are the same as
those for Fig. \ref{fig2}. For the locations of the peaks in the region $%
0<\theta <\pi /2$ one has $\theta \approx 1.032$ for $v/c=0.9$ and $\theta
\approx 0.807$ for $v/c=0.99$. For the roots of the equation $\alpha
_{n}^{(0)}=0$ one gets $\theta \approx 1.033$ for $v/c=0.9$ and $\theta
\approx 0.808$ for $v/c=0.99$. Again we have a good agreement with the
analytical estimates of the locations of the peaks.

\begin{figure}[tbph]
\begin{center}
\epsfig{figure=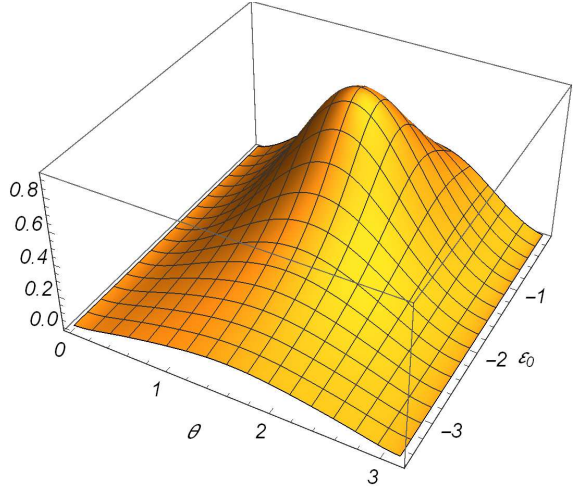,width=7.5cm,height=6.5cm}
\end{center}
\caption{The angular density of the number of the radiated quanta as a
function of $\protect\theta $ and of cylinder dielectric permittivity $%
\protect\varepsilon _{0}$. For the remaining parameters we have taken $%
v/c=0.8$, $r_{c}/r_{q}=0.95$, $n=1$.}
\label{fig5}
\end{figure}

\begin{figure}[tbph]
\begin{center}
\epsfig{figure=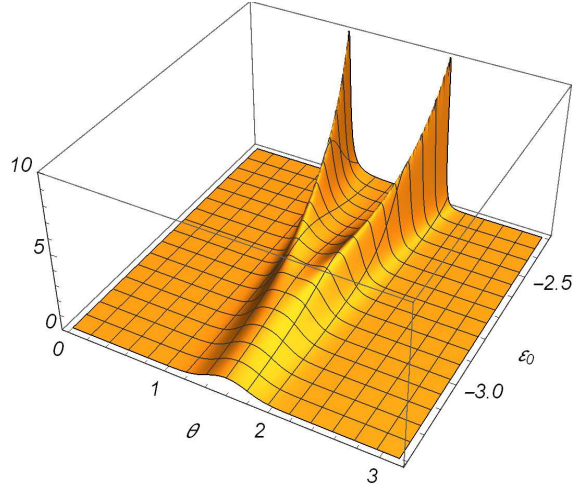,width=7.5cm,height=6.5cm}
\end{center}
\caption{The same as in Fig. \protect\ref{fig5} for $n=10$.}
\label{fig6}
\end{figure}

In order to see the dependence of the radiation intensity on the dielectric
permittivity of the cylinder, in Figs. \ref{fig5} and \ref{fig6} the
quantity $(\hbar c/q^{2})dN_{n}/d\Omega $ is plotted as a function of $%
\theta $ and $\varepsilon _{0}$ for $r_{c}/r_{q}=0.95$, $v/c=0.8$, $n=1$
(Fig. \ref{fig5}) and $n=10$ (Fig. \ref{fig6}). In Fig. \ref{fig6} we see
the formation of angular peaks with increasing $\varepsilon _{0}$. For the
peak $\theta =\theta _{p}$, $0<\theta _{p}<\pi /2$, the angle $\theta _{p}$
and the width decrease with increasing $\varepsilon _{0}$, whereas the
corresponding height increases. This behavior is in agreement with
analytical estimates given before.

\begin{figure}[tbph]
\begin{center}
\epsfig{figure=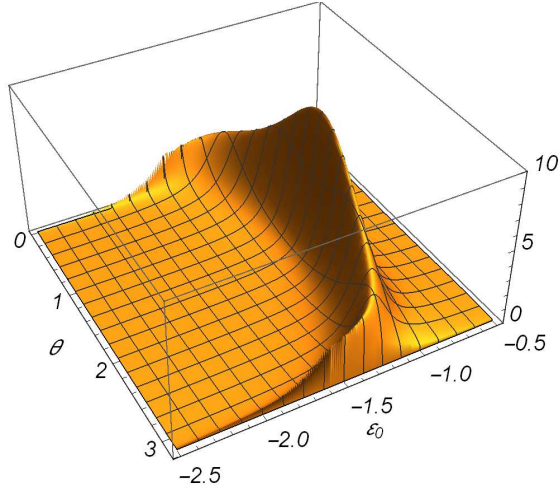,width=7.5cm,height=6.5cm}
\end{center}
\caption{The same as in Fig. \protect\ref{fig5} for $v/c=0.2$ and $n=1$.}
\label{fig7}
\end{figure}

In the discussion above we have argued that for a cylinder with negative
dielectric permittivity the behavior of the radiation intensity in the
nonrelativistic limit can be essentially different from that in the case of
positive permittivity if $\varepsilon _{0}^{\prime }$ is sufficiently close
to $-\varepsilon _{1}$. This feature is seen from Fig. \ref{fig7}, where we
have plotted the angular density of the number of the radiated quanta (in
units of $q^{2}/(\hbar c)$) as a function of $\theta $ and $\varepsilon _{0}$
for $r_{c}/r_{q}=0.95$, $v/c=0.2$, $n=1$.

\begin{figure}[tbph]
\begin{center}
\epsfig{figure=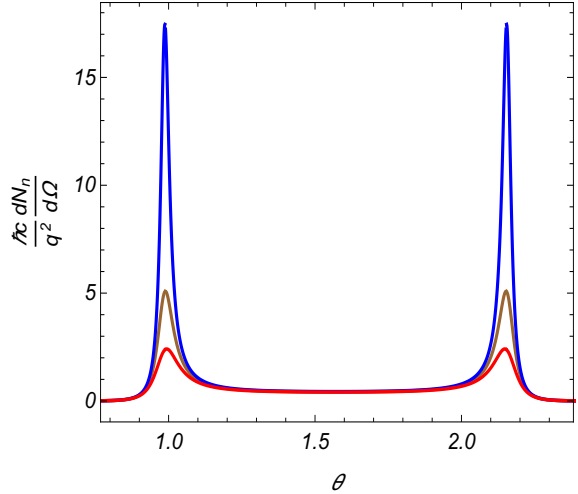,width=7.5cm,height=6.5cm}
\end{center}
\caption{The angular density of the number of the radiated quanta as a
function of $\protect\theta $ for different values of the imaginary part of
dielectric permittiviy. For the values of the parameters see the text.}
\label{fig8}
\end{figure}

As it has been mentioned above, the height of the peaks in the angular
distribution of the radiation intensity is increasing with increasing
harmonic number. However, it should be noted that in realistic physical
situation this increase is restricted by several factors. One of them may be
taking into account the imaginary part of the dielectric permittivity, $%
\varepsilon _{0}=\varepsilon _{0}^{\prime }+i\varepsilon _{0}^{\prime \prime
}$. In Fig. \ref{fig8} we display the angular density $(\hbar
c/q^{2})dN_{n}/d\Omega $ as a function of $\theta $ for $r_{c}/r_{q}=0.95$, $%
v/c=0.8$, $n=10$, $\varepsilon _{0}^{\prime }=-2.2$ and for the ratios $%
\varepsilon _{0}^{\prime \prime }/\varepsilon _{0}^{\prime }=0,0.01,0.02$
(with decreasing heights of the peaks). As it is seen from the graphs, the
influence of the imaginary part is essential only near the peaks.

\begin{figure}[tbph]
\begin{center}
\epsfig{figure=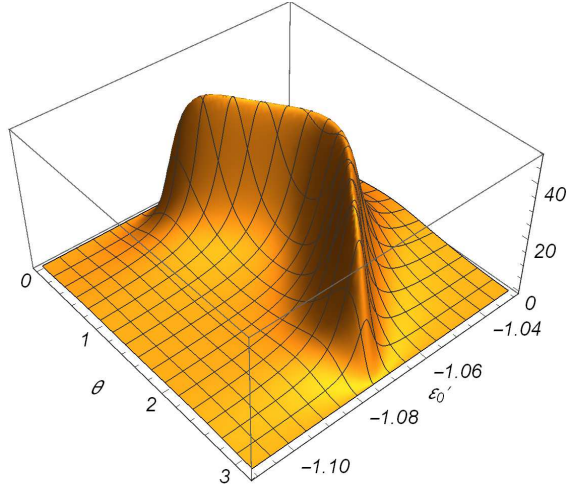,width=7.5cm,height=6.5cm}
\end{center}
\caption{The dependence of the angular density of the number of the radiated
quanta on $\protect\theta $ and on the real part of dielectric permittivity
for $n=2$ and $\protect\varepsilon _{0}^{\prime \prime }/\protect\varepsilon %
_{0}^{\prime }=0.0025$. The values of the remaining parameters coincide with
those for Fig. \protect\ref{fig7}.}
\label{fig9}
\end{figure}

In Fig. \ref{fig9} the angualr density of the radiated quanta number is
plotted as a function of $\theta $ and of the real part of the dielectric
permittivity $\varepsilon _{0}^{\prime }$ for the harmonic $n=2$ and for $%
\varepsilon _{0}^{\prime \prime }/\varepsilon _{0}^{\prime }=0.0025$. The
values of the other parameters are the same as those for Fig. \ref{fig7}.
Again, we see an essential amplification of the radiation intensity in the
region for the values of $\varepsilon _{0}^{\prime }$ close to $-\varepsilon
_{1}$. Compared to the case depicted in Fig. \ref{fig7}, here the region is
narrower.

\begin{figure}[tbph]
\begin{center}
\epsfig{figure=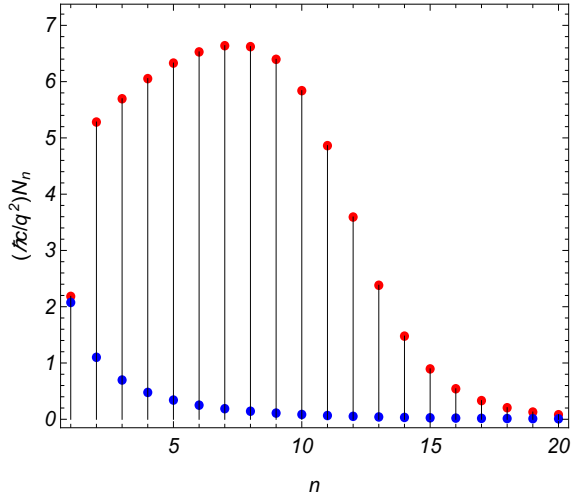,width=7.5cm,height=6.5cm}
\end{center}
\caption{The total number of the radiated quanta as a function of the
harmonic $n$ for $v/c=0.8$. The remaining parameters are the same as those
in Fig. \protect\ref{fig2}.}
\label{fig10}
\end{figure}

\begin{figure}[tbph]
\begin{center}
\epsfig{figure=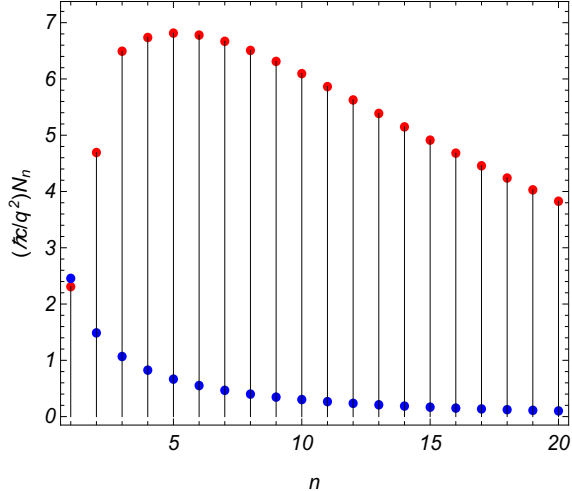,width=7.5cm,height=6.5cm}
\end{center}
\caption{The same as in Fig. \protect\ref{fig8} for $v/c=0.9$.}
\label{fig11}
\end{figure}

It is also of interest to see the influence of the cylinder on the total
radiated energy on a given harmonic, obtained from Eq. (\ref{In4}) by
integration over the angles $\theta $ and $\phi $. Figure \ref{fig10}
presents the total number of the radiated quanta per period of charge
rotation, $N_{n}=\int d\Omega \,(dN_{n}/d\Omega )$ (red circles), versus the
number of the harmonic $n$, for $v/c=0.8$ and for the same values of the
parameters as in Fig. \ref{fig2}. The blue points present the same quantity
for the radiation in the vacuum. The corresponding data for $v/c=0.9$ are
displayed in Fig. \ref{fig11}.

As it is seen from the numerical data in Figs. \ref{fig10} and \ref{fig11},
the radiation intensity on a given harmonic, integrated over the angles, can
be essentially amplified by the presence of the cylinder. For large
harmonics, $n\gg 1/\zeta (v/c)$ (with the function $\zeta (y)$ from Eq. (\ref%
{dset})) the radiation intensity tends to the one for the radiation in
vacuum. Physically, the latter feature is related to the fact that for those
harmonics the Fourier components of the charge field are strongly localized
near the particle trajectory and the corresponding polarization of the
cylinder is week. The function $\zeta (y)$ is monotonically decreasing and
with increasing $n$ the approaching the radiation intensity to the one in
the vacuum is slower for larger $v/c$.

\section{Summary}

\label{sec:Conc}

We have investigated the features of the synchrotron radiation from a
charged particle rotating around a cylinder in the spectral range where the
real part of the dielectric permittivity is negative. The importance of the
investigation of electrodynamical effects in that range is partially
motivated by potential applications in plasmonics. In the problem under
consideration two types of radiations are present. The first one corresponds
to the surface-type modes (surface polaritons) which are localized near the
cylinder surface. The corresponding radiation intensity was considered in
Ref. \cite{Kota19}. The second type of radiation corresponds to the waves
propagating at large distances from the cylinder. The corresponding angular
density of the radiation intensity on a given harmonic is given by Eq. (\ref%
{In4}). Compared to the case of a cylinder with positive dielectric
permittivity, a qualitatively new feature appears in the nonrelativistic
limit when the real part of the permittivity is sufficiently close to $%
-\varepsilon _{1}$. In this region the angular density of the radiation
intensity behaves as $\beta _{1}^{2(n-1)}$ for radiation harmonics $n\geq 2$
and like $1/\ln ^{2}(\beta _{1})$ for $n=1$. For $|\varepsilon _{0}^{\prime
}+\varepsilon _{1}|\gg \beta _{1}^{2}$ the angular density decays as $\beta
_{1}^{2n+2}$ for $n\geq 1$. Similar differences from the case of a cylinder
with positive dielectric permittivity arise in the behavior of the radiation
intensity for small values of the angle $\theta $.

An interesting feature in the influence of the cylinder on the radiation at
large distances from the cylinder is the possibility for the appearance of
strong narrow peaks in the angular distribution of the radiation intensity
on large harmonics. They are located in the angular region $\sin \theta
<1/\beta _{c}$, with $\beta _{c}$ being the velocity of the charge image on
the cylinder surface. We gave analytic estimates for the location, height
and the width of the peaks. With decreasing values of the angle $\theta $ in
the range $0<\theta <\pi /2$ the height of the peak increases, whereas the
width decreases. Similar behavior takes place with increasing number of the
radiation harmonic. Among the factors that limits the increase of the peak
height is the imaginary part of the cylinder dielectric permittivity. The
presence of the cylinder may lead also to an essential increase of the
integrated intensity of the radiation. All these features we have
demonstrated on numerical examples.

\section*{Acknowledgement}

A.A.S., L.Sh.G. and H.F.Kh. were supported by Grant No.~18T-1C397 from the
Science Committee of the Ministry of Education and Science of the Republic
of Armenia.

\end{document}